\pdfoutput=1

\documentclass[11pt]{article}

\usepackage[]{ACL2023}

\usepackage{times}
\usepackage{latexsym}

\usepackage[T1]{fontenc}

\usepackage[utf8]{inputenc}
\usepackage[hang,flushmargin]{footmisc}

\usepackage{microtype}

\usepackage{inconsolata}

\usepackage{booktabs}       
\usepackage{url}            
\usepackage{amsfonts}       
\usepackage{amsmath}
\usepackage{graphicx}
\usepackage{multirow}
\usepackage{pifont}
\usepackage{natbib}
\newcommand{\cmark}{\ding{51}}%
\newcommand{\xmark}{\ding{55}}%

\newcommand\our{\textsc{SimLM}}

\title{\our{}: Pre-training with Representation Bottleneck for \\ Dense Passage Retrieval}

\author{Liang Wang,~Nan Yang,~Xiaolong Huang,~Binxing Jiao\\
~\textbf{Linjun Yang},~\textbf{Daxin Jiang},~\textbf{Rangan Majumder},~\textbf{Furu Wei}\\
Microsoft Corporation \\
\{wangliang,nanya,xiaolhu,binxjia,yang.linjun,djiang,ranganm,fuwei\}@microsoft.com \\}

\begin{document}
\maketitle
\begin{abstract}
In this paper,
we propose \our{} (\textbf{Sim}ilarity matching with \textbf{L}anguage \textbf{M}odel pre-training),
a simple yet effective pre-training method for dense passage retrieval.
It employs a simple bottleneck architecture that learns to compress the passage information
into a dense vector through self-supervised pre-training.
We use a replaced language modeling objective, which is inspired by ELECTRA~\citep{Clark2020ELECTRAPT},
to improve the sample efficiency and reduce the mismatch of the input distribution between pre-training and fine-tuning.
\our{} only requires access to an unlabeled corpus
and is more broadly applicable when there are no labeled data or queries.
We conduct experiments on several large-scale passage retrieval datasets
and show substantial improvements over strong baselines under various settings.
Remarkably,
\our{} even outperforms multi-vector approaches such as ColBERTv2 ~\citep{Santhanam2021ColBERTv2EA}
which incurs significantly more storage cost.
Our code and model checkpoints are available at ~\url{https://github.com/microsoft/unilm/tree/master/simlm}.
\end{abstract}

\section{Introduction}

Passage retrieval is an important component in applications
like ad-hoc information retrieval,
open-domain question answering ~\citep{Karpukhin2020DensePR},
retrieval-augmented generation ~\citep{Lewis2020RetrievalAugmentedGF}
and fact verification ~\citep{Thorne2018TheFE}.
Sparse retrieval methods such as BM25 were the dominant approach for several decades,
and still play a vital role nowadays.
With the emergence of large-scale pre-trained language models (PLM) ~\citep{Devlin2019BERTPO},
increasing attention is being paid to neural dense retrieval methods ~\citep{Lin2021PretrainedTF}.
Dense retrieval methods map both queries and passages into a low-dimensional vector space,
where the relevance between the queries and passages are measured by the dot product or cosine similarity between their respective vectors.

\begin{table}[ht]
\centering
\scalebox{0.9}{\begin{tabular}{@{}ccc@{}}
\hline
PLM & MS-MARCO & GLUE \\ \hline
BERT   &  \textbf{33.7} &  80.5  \\
RoBERTa  & 33.1  &  88.1  \\
ELECTRA  &  31.9  &  \textbf{89.4}  \\ \hline
\end{tabular}}
\caption{Inconsistent performance trends between different models on retrieval task and NLU tasks.
We report MRR@10 on the dev set of MS-MARCO passage ranking dataset
and test set results on GLUE benchmark.
Details are available in the Appendix ~\ref{sec:inconsistent}.}
\label{tab:inconsistent_perf}
\end{table}

Like other NLP tasks, dense retrieval benefits greatly from a strong general-purpose pre-trained language model.
However,
general-purpose pre-training does not solve all the problems.
As shown in Table ~\ref{tab:inconsistent_perf},
improved pre-training techniques that are verified by benchmarks like GLUE ~\citep{Wang2018GLUEAM}
do not result in consistent performance gain for retrieval tasks.
Similar observations are also made by ~\citet{Lu2021LessIM}.
We hypothesize that,
to perform robust retrieval,
the [CLS] vector used for computing matching scores
should encode all the essential information in the passage.
The next-sentence prediction (NSP) task in BERT
introduces some supervision signals for the [CLS] token,
while RoBERTa ~\citep{Liu2019RoBERTaAR} and ELECTRA do not have such sequence-level tasks.

In this paper,
we propose SimLM to pre-train a representation bottleneck with replaced language modeling objective.
SimLM consists of a deep encoder and a shallow decoder
connected with a representation bottleneck,
which is the [CLS] vector in our implementation.
Given a randomly masked text segment,
we first employ a generator to sample replaced tokens for masked positions,
then use both the deep encoder and shallow decoder to predict the original tokens at \emph{all} positions.
Since the decoder only has limited modeling capacity,
it must rely on the representation bottleneck to perform well on this pre-training task.
As a result,
the encoder will learn to compress important semantic information into the bottleneck,
which would help train biencoder-based ~\footnote{Also called dual-encoder / two-tower encoder.} dense retrievers.
Our pre-training objective works with plain texts and
does not require any generated pseudo-queries as for GPL ~\citep{wang2022gpl}.

Compared to existing pre-training approaches such as Condenser ~\citep{Gao2021CondenserAP} or coCondenser ~\citep{Gao2022UnsupervisedCA},
our method has several advantages.
First,
it does not have any extra skip connection between the encoder and decoder,
thus reducing the bypassing effects and simplifying the architecture design.
Second,
similar to ELECTRA pre-training,
our replaced language modeling objective can back-propagate gradients at \emph{all} positions
and does not have [MASK] tokens in the inputs during pre-training.
Such a design increases sample efficiency and
decreases the input distribution mismatch between pre-training and fine-tuning.

To verify the effectiveness of our method,
we conduct experiments on several large-scale web search and open-domain QA datasets:
MS-MARCO passage ranking ~\citep{Campos2016MSMA},
TREC Deep Learning Track datasets,
and the Natural Questions (NQ) dataset ~\citep{Kwiatkowski2019NaturalQA}.
Results show substantial gains over other competitive methods using BM25 hard negatives only.
When combined with mined hard negatives and cross-encoder based re-ranker distillation,
we can achieve new state-of-the-art performance.

\section{Related Work}

\noindent
\textbf{Dense Retrieval }
The field of information retrieval (IR) ~\citep{Manning2005IntroductionTI}
aims to find the relevant information given an ad-hoc query
and has played a key role in the success of modern search engines.
In recent years,
IR has witnessed a paradigm shift from traditional BM25-based inverted index retrieval
to neural dense retrieval ~\citep{Lin2021PretrainedTF,Karpukhin2020DensePR}.
BM25-based retrieval, though efficient and interpretable,
suffers from the issue of lexical mismatch between the query and passages.
Methods like document expansion ~\citep{Nogueira2019DocumentEB} or query expansion ~\citep{Azad2019QueryET,Wang2023Query2docQE}
are proposed to help mitigate this issue.
In contrast,
neural dense retrievers first map the query and passages to a low-dimensional vector space,
and then perform semantic matching.
Popular methods include DSSM ~\citep{Huang2013LearningDS}, C-DSSM ~\citep{Shen2014LearningSR},
and DPR ~\citep{Karpukhin2020DensePR} etc.
Inference can be done efficiently with approximate nearest neighbor (ANN) search algorithms
such as HNSW ~\citep{Malkov2020EfficientAR}.

Some recent works ~\citep{Chen2021SalientPA, Reimers2021TheCO, Sciavolino2021SimpleEQ} show that
neural dense retrievers may fail to capture some exact lexical match information.
To mitigate this issue,
~\citet{Chen2021SalientPA} proposes to use BM25 as a complementary teacher model,
ColBERT ~\citep{Khattab2020ColBERTEA} instead replaces simple dot-product matching
with a more complex token-level MaxSim interaction,
while COIL ~\citep{Gao2021COILRE} incorporates lexical match information
into the scoring component of neural retrievers.
Our proposed pre-training method aims to adapt the underlying text encoders for retrieval tasks,
and can be easily integrated with existing approaches.
\newline

\noindent
\textbf{Pre-training for Dense Retrieval }
With the development of large-scale language model pre-training ~\citep{Dong2019UnifiedLM, Clark2020ELECTRAPT},
Transformer-based models such as BERT ~\citep{Devlin2019BERTPO} have become
the de facto backbone architecture for learning text representations.
However,
most pre-training tasks are designed without any prior knowledge of downstream applications.
~\citet{Chang2020PretrainingTF} presents three heuristically constructed pre-training tasks tailored for text retrieval:
inverse cloze task (ICT), body first selection (BFS), and wiki link prediction (WLP).
These tasks exploit the document structure of Wikipedia pages
to automatically generate contrastive pairs.
Other related pre-training tasks include representative words prediction ~\citep{Ma2021PROPPW},
contrastive span prediction ~\citep{Ma2022PretrainAD},
contrastive learning with independent cropping ~\citep{Izacard2021UnsupervisedDI},
domain-matched pre-training ~\citep{oguz2022domain}
or neighboring text pairs ~\citep{Neelakantan2022TextAC} etc.

\begin{figure*}[ht]
\begin{center}
 \includegraphics[width=1.0\linewidth]{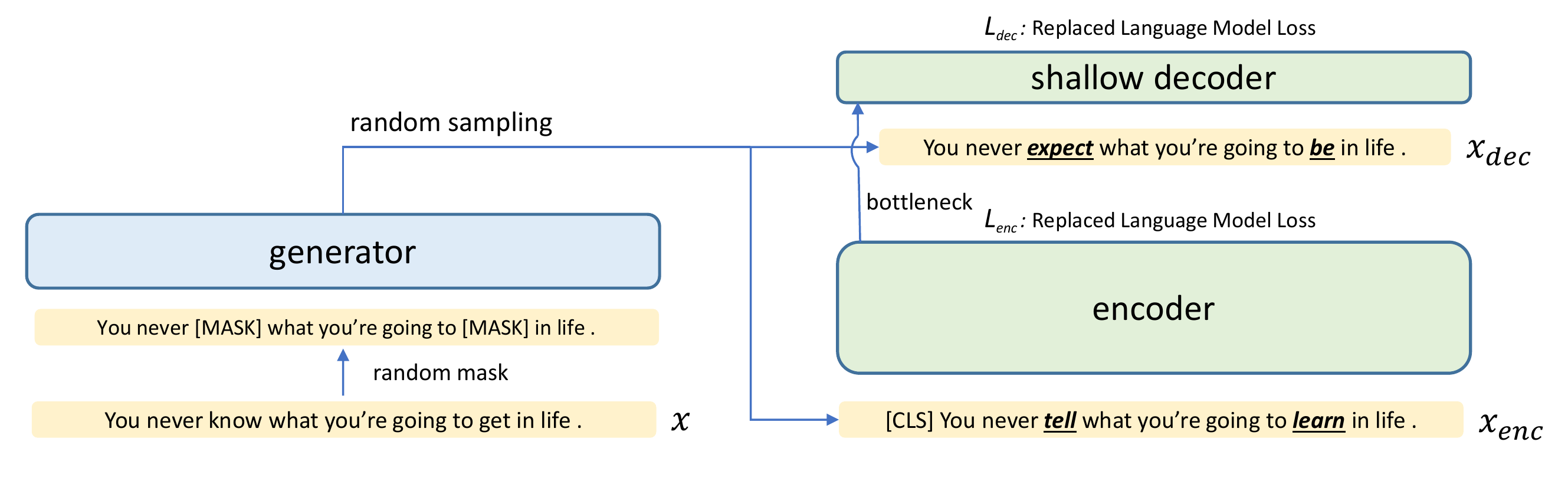}
 \caption{Pre-training architecture of SimLM.
 Replaced tokens (underlined) are randomly sampled from the generator distribution.}
 \label{fig:pretrain}
\end{center}
\end{figure*}

Another line of research builds upon the intuition
that the [CLS] vector should encode all the important information in the given text for robust matching,
which is also one major motivation for this paper.
Such methods include Condenser ~\citep{Gao2021CondenserAP}, coCondenser ~\citep{Gao2022UnsupervisedCA},
SEED ~\citep{Lu2021LessIM}, DiffCSE ~\citep{Chuang2022DiffCSEDC}, and RetroMAE ~\citep{Liu2022RetroMAEPR} etc.
Compared with Condenser and coCondenser,
our pre-training architecture does not have skip connections between the encoder and decoder,
and therefore forces the [CLS] vector to encode as much information as possible.
RetroMAE ~\citep{Liu2022RetroMAEPR} is a concurrent work at the time of writing that
combines a bottleneck architecture and the masked auto-encoding objective.

\section{SimLM}

\subsection{Pre-training}

For pre-training,
we assume there is a collection of passages $\mathbb{C} = \{\mathbf{x}_i\}_{i=1}^{|\mathbb{C}|}$,
where $\mathbf{x}$ denotes a single passage.
Since our motivation is to have a general pre-training method,
we do not assume access to any query or human-labeled data.

The overall pre-training architecture is shown in Figure ~\ref{fig:pretrain}.
Given a text sequence $\mathbf{x}$,
its tokens are randomly replaced with probability $p$ by two sequential operations:
random masking with probability $p$ denoted as
$\mathbf{x}' = \text{Mask(}\mathbf{x}, p\text{)}$,
and then sampling from an ELECTRA-style generator $g$
denoted as $\text{Sample(}g, \mathbf{x}'\text{)}$.
Due to the randomness of sampling,
a replaced token can be the same as the original one.
The above operations are performed twice with potentially
different replace probabilities $p_\text{enc}$ and $p_\text{dec}$
to get the encoder input $\mathbf{x}_\text{enc}$ and decoder input $\mathbf{x}_\text{dec}$.
\begin{equation}
\begin{aligned}
    \mathbf{x}_{\text{enc}} & = \text{Sample(}g,\ \text{Mask(}\mathbf{x},\ p_{\text{enc}}\text{))} \\
    \mathbf{x}_{\text{dec}} & = \text{Sample(}g,\ \text{Mask(}\mathbf{x},\ p_{\text{dec}}\text{))}
\end{aligned}
\end{equation}
We also make sure that any replaced token
in $\mathbf{x}_\text{enc}$ is also replaced in $\mathbf{x}_\text{dec}$
to increase the difficulty of the pre-training task.

The encoder is a deep multi-layer Transformer that can be initialized with pre-trained models like BERT ~\citep{Devlin2019BERTPO}.
It takes $\mathbf{x}_\text{enc}$ as input
and outputs the last layer [CLS] vector $\mathbf{h}_\text{cls}$ as a representation bottleneck.
The decoder is a 2-layer shallow Transformer with a language modeling head
and takes $\mathbf{x}_\text{dec}$ and $\mathbf{h}_\text{cls}$ as inputs.
Unlike the decoder component in autoregressive sequence-to-sequence models,
the self-attention in our decoder is bi-directional.
The pre-training task is replaced language modeling for both the encoder and decoder,
which predicts the tokens before replacement at \emph{all} positions.
The loss function is the token-level cross-entropy.
The encoder loss $L_\text{enc}$ is shown as follows:
\begin{equation}
    \min\ \ L_{\text{enc}} = -\frac{1}{|\mathbf{x}|}\sum_{i=1}^{|\mathbf{x}|} \log p(\mathbf{x}[i]\ |\ \mathbf{x}_\text{enc})
\end{equation}
Similarly for the decoder loss $L_\text{dec}$.
The final pre-training loss is their simple sum: $L_\text{pt} = L_\text{enc} + L_\text{dec}$.
We do not fine-tune the parameters of the generator
as our preliminary experiments do not show any performance gain.

It is often reasonable to assume access to the target retrieval corpus
before seeing any query.
Therefore,
we directly pre-train on the target corpus similar to coCondenser ~\citep{Gao2022UnsupervisedCA}.
After the pre-training finishes,
we throw away the decoder and
only keep the encoder for supervised fine-tuning.

Since the decoder has very limited modeling capacity,
it needs to rely on the representation bottleneck to perform well on the pre-training task.
For the encoder,
it should learn to compress all the semantic information and pass it to the decoder through the bottleneck.

\subsection{Fine-tuning}

\begin{figure*}[ht]
\begin{center}
 \includegraphics[width=1.0\linewidth]{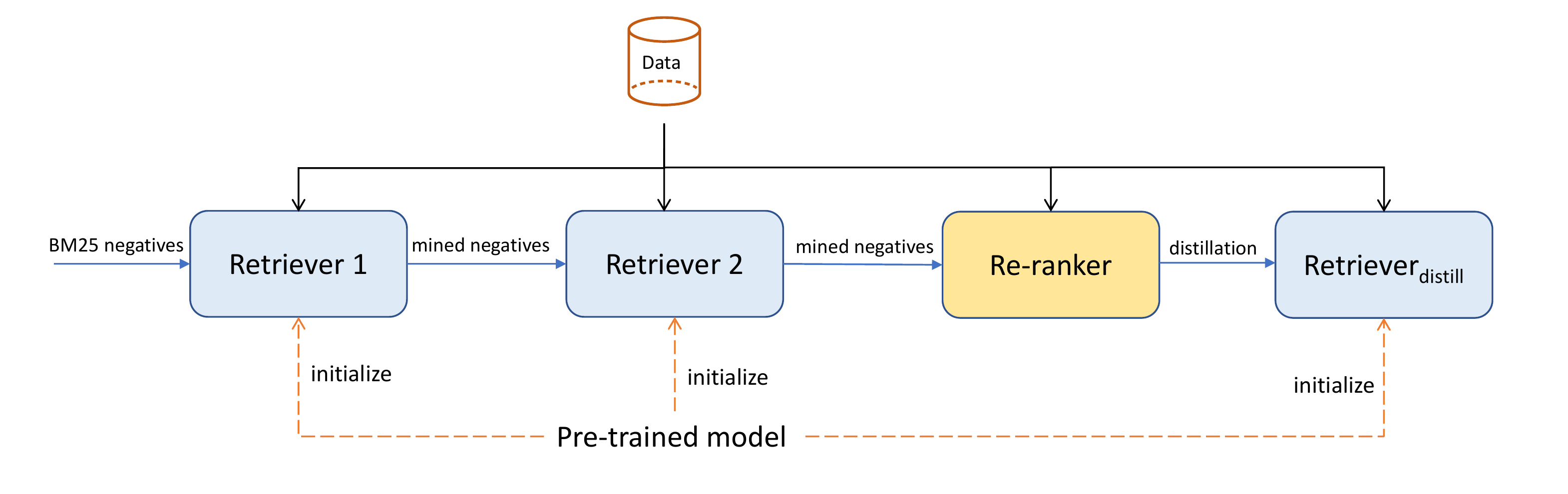}
 \caption{Illustration of our supervised fine-tuning pipeline.
 Note that we only use SimLM to initialize the biencoder-based retrievers.
 For cross-encoder based re-ranker,
 we use off-the-shelf pre-trained models such as ELECTRA$_\text{base}$.}
 \label{fig:finetune}
\end{center}
\end{figure*}

Compared to training text classification or generation models,
training state-of-the-art dense retrieval models requires a relatively complicated procedure.
In Figure ~\ref{fig:finetune},
we show our supervised fine-tuning pipeline.
In contrast to previous approaches,
our proposed pipeline is relatively straightforward
and does not require joint training ~\citep{Ren2021RocketQAv2AJ}
or re-building index periodically ~\citep{Xiong2021ApproximateNN}.
Each stage takes the outputs from the previous stage as inputs
and can be trained in a standalone fashion.
\newline

\noindent
\textbf{Retriever$_\text{1}$ }
Given a labeled query-passage pair ($q^+, d^+$),
we take the last-layer [CLS] vector of the pre-trained encoder
as their representations ($\mathbf{h}_{q^+}, \mathbf{h}_{d^+}$).
Both the in-batch negatives and BM25 hard negatives
are used to compute the contrastive loss $L_\text{cont}$:
\begin{equation} \label{equ:infonce}
    -\log \frac{\phi(q^+, d^+)}{\phi(q^+, d^+) + \displaystyle\sum_{n_i \in \mathbb{N}}(\phi(q^+, n_i) + \phi(d^+, n_i))}
\end{equation}
Where $\mathbb{N}$ denotes all the negatives,
and $\phi(q, d)$ is a function to compute the matching score between query $q$ and passage $d$.
In this paper,
we use temperature-scaled cosine similarity function:
$\phi(q, d) = \text{exp}(\frac{1}{\tau}\cos(\mathbf{h}_q, \mathbf{h}_d))$.
$\tau$ is a temperature hyper-parameter and
set to a constant $0.02$ in our experiments.
\newline

\noindent
\textbf{Retriever$_\text{2}$ }
It is trained in the same way as Retriever$_\text{1}$ except that
the hard negatives are mined based on a well-trained Retriever$_\text{1}$ checkpoint.
\newline

\noindent
\textbf{Re-ranker }
is a cross-encoder that re-ranks the top-$k$ results of Retriever$_\text{2}$.
It takes the concatenation of query $q$ and passage $d$ as input
and outputs a real-valued score $\theta(q, d)$.
Given a labeled positive pair ($q^+, d^+$)
and $n - 1$ hard negative passages randomly sampled
from top-$k$ predictions of Retriever$_\text{2}$,
we adopt a listwise loss to train the re-ranker:
\begin{equation}
    -\log \frac{\text{exp}(\theta(q^+, d^+))}{\text{exp}(\theta(q^+, d^+)) + \sum_{i=1}^{n-1}\text{exp}(\theta(q^+, d_{i}^{-}))}
\end{equation}

The cross-encoder architecture can model the full interaction between the query and the passage,
making it suitable to be a teacher model for knowledge distillation.
\newline

\noindent
\textbf{Retriever$_\text{distill}$ }
Although cross-encoder based re-ranker is powerful,
it is not scalable enough for first-stage retrieval.
To combine the scalability of biencoder and the effectiveness of cross-encoder,
we can train a biencoder-based retriever by distilling the knowledge from the re-ranker.
The re-ranker from the previous stage is employed to
compute scores for both positive pairs and mined negatives from Retriever$_\text{2}$.
These scores are then used as training data for knowledge distillation.
With $n - 1$ mined hard negatives,
we use KL (Kullback-Leibler) divergence $L_\text{kl}$
as the loss function for distilling the soft labels:
\begin{equation}
    L_\text{kl} = \sum_{i=1}^{n} p_\text{ranker}^i \log \frac{p_\text{ranker}^i}{p_\text{ret}^i}
\end{equation}
where $p_\text{ranker}$ and $p_\text{ret}$ are normalized probabilities from
the re-ranker teacher and Retriever$_\text{distill}$ student.
For training with the hard labels,
we use the contrastive loss $L_\text{cont}$ as defined in Equation ~\ref{equ:infonce}.
The final loss is their linear interpolation: $L = L_\text{kl} + \alpha L_\text{cont}$.

\begin{table*}[ht]
\centering
\scalebox{0.83}{\begin{tabular}{@{}lccccccc@{}}
\hline
\multirow{2}{*}{Model} & \multirow{2}{*}{+distill} & \begin{tabular}[c]{@{}c@{}}\\ single\\ vector?\end{tabular} & \multicolumn{3}{c}{MS MARCO dev} & TREC DL 19 & TREC DL 20 \\
              &     &    & MRR@10    & R@50    & R@1k    & nDCG@10    & nDCG@10    \\ \hline
\textbf{Sparse retrieval} & & &  &  &   &  &   \\
BM25   &  & \cmark & 18.5 &  58.5   &  85.7   &   51.2$^*$  &  47.7$^*$   \\
DeepCT ~\citep{Dai2019ContextAwareST} &  & \cmark &  24.3  & 69.0 &  91.0   &  57.2   &   -   \\
docT5query ~\citep{nogueira2019doc2query} &  & \cmark  &  27.7  &  75.6  & 94.7  &  64.2  &  -  \\ \hline
\textbf{Dense retrieval} & & &  &  &   &  &   \\
ANCE ~\citep{Xiong2021ApproximateNN} &  & \cmark  &  33.0 & -  &  95.9  &   64.5$^\dagger$  & 64.6$^\dagger$ \\
SEED ~\citep{Lu2021LessIM} &  & \cmark  & 33.9 &  -  &  96.1  &   -  & -  \\
TAS-B ~\citep{Hofsttter2021EfficientlyTA} & \cmark & \cmark &  34.0 &  -  &  97.5  &  71.2  & 69.3  \\
RetroMAE ~\citep{Liu2022RetroMAEPR} &  & \cmark &  35.0 &  -  &  97.6  &  -  & -  \\
COIL ~\citep{Gao2021COILRE} &  & &  35.5 &  -  &  96.3  &  70.4  & -  \\
ColBERT ~\citep{Khattab2020ColBERTEA} &  &  &   36.0  &   82.9  &  96.8   &  -  & - \\
Condenser ~\citep{Gao2021CondenserAP} &  & \cmark  &  36.6  &   -   &  97.4   &  69.8   &  -  \\
RocketQA ~\citep{Qu2021RocketQAAO}  & \cmark & \cmark &  37.0   &   85.5  &   97.9   &  -  & -  \\
PAIR ~\citep{Ren2021PAIRLP} & \cmark & \cmark & 37.9  &  86.4  &  98.2   &  -  & -  \\
coCondenser ~\citep{Gao2022UnsupervisedCA} & & \cmark &  38.2   &  86.5$^*$   &  98.4  &  \textbf{71.7}$^*$  &  68.4$^*$  \\
RocketQAv2 ~\citep{Ren2021RocketQAv2AJ} &  \cmark & \cmark &  38.8  &  86.2    &   98.1   &  - &  - \\
AR2 ~\citep{zhang2021adversarial} &  \cmark  &  \cmark &   39.5  &  \textbf{87.8}  &  98.6   &  - &  - \\
ColBERTv2 ~\citep{Santhanam2021ColBERTv2EA} &  \cmark  &  &   39.7  &  86.8    &   98.4   &  - &  - \\ \hline
\our{}   &  \cmark  & \cmark  &   \textbf{41.1}  &  \textbf{87.8}  &  \textbf{98.7}  &  71.4  &  \textbf{69.7}  \\ \hline
\end{tabular}}
\caption{Main results on MS-MARCO passage ranking and TREC datasets.
Results with * are from our reproduction with public checkpoints.
$\dagger$: from Pyserini ~\citep{Lin2021PyseriniAP}.}
\label{tab:ir_results}
\end{table*}

Our pre-trained SimLM model is used to initialize all three biencoder-based retrievers
but not the cross-encoder re-ranker.
Since our pre-training method only affects model initialization,
it can be easily integrated into other more effective training pipelines.

\section{Experiments}

\subsection{Setup}

\noindent
\textbf{Datasets and Evaluation }
We use MS-MARCO passage ranking ~\citep{Campos2016MSMA},
TREC Deep Learning (DL) Track 2019 ~\citep{craswell2020overview}
and 2020 ~\citep{Craswell2020OverviewOT},
Natural Questions (NQ) ~\citep{Kwiatkowski2019NaturalQA, Karpukhin2020DensePR} datasets for training and evaluation.
The MS-MARCO dataset is based on Bing search results
and consists of about $500k$ labeled queries and $8.8M$ passages.
Since the test set labels are not publicly available,
we report results on the development set with $6980$ queries.
The NQ dataset is targeted for open QA
with about $80k$ question-answer pairs in the training set
and $21M$ Wikipedia passages.
For evaluation metrics,
we use MRR@10, Recall@50, and Recall@1k for MS-MARCO,
nDCG@10 for TREC DL,
and Recall@20, Recall@100 for the NQ dataset.
\newline

\noindent
\textbf{Implementation Details } ~\label{sec:impl_details}
For pre-training,
we initialize the encoder with BERT$_\text{base}$ (uncased version).
The decoder is a two-layer Transformer
whose parameters are initialized with the last two layers of BERT$_\text{base}$.
The generator is borrowed from the ELECTRA$_\text{base}$ generator,
and its parameters are frozen during pre-training.
We pre-train for $80k$ steps for MS-MARCO corpus
and $200k$ steps for NQ corpus,
which roughly correspond to $20$ epochs.
Pre-training is based on $8$ V100 GPUs.
With automatic mixed-precision training,
it takes about $1.5$ days and $3$ days for the MS-MARCO and NQ corpus respectively.

For more implementation details,
please check out the Appendix section ~\ref{sec:more_impl_details}.

\subsection{Main Results}

We list the main results in Table ~\ref{tab:ir_results} and ~\ref{tab:dpr_result}.
For the MS-MARCO passage ranking dataset,
the numbers are based on the Retriever$_\text{distill}$ in Figure ~\ref{fig:finetune}.
Our method establishes new state-of-the-art with MRR@10 41.1,
even outperforming multi-vector methods like ColBERTv2.
As shown in Table ~\ref{tab:compare_colbert},
ColBERTv2 has a 6x storage cost as
it stores one vector per token instead of one vector per passage.
It also requires a customized two-stage index search algorithm during inference,
while our method can utilize readily available vector search libraries.

The TREC DL datasets have more fine-grained human annotations,
but also much fewer queries (less than 100 labeled queries).
We find that using different random seeds could have a 1\%-2\% difference in terms of nDCG@10.
Though our model performs slightly worse on the 2019 split compared to coCondenser,
we do not consider such difference as significant.

\begin{table}[ht]
\centering
\scalebox{0.95}{\begin{tabular}{lcc}
\hline
          & Index size & Index search \\ \hline
ColBERTv2 & \textgreater 150GB      & Two-stage  \\
\our{}      & 27GB       & One-stage  \\ \hline
\end{tabular}}
\caption{Comparison with ColBERTv2 ~\citep{Santhanam2021ColBERTv2EA}
in terms of index storage cost (w/o any compression)
and complexity of index search algorithms.}
\label{tab:compare_colbert}
\end{table}

\begin{table}[ht]
\centering
\scalebox{0.88}{\begin{tabular}{@{}lcc@{}}
\toprule
\multirow{2}{*}{Model} & \multicolumn{2}{c}{NQ} \\
                       & R@20      & R@100      \\ \hline
BM25  &   59.1   &   73.7 \\
DPR$_\text{single}$ ~\citep{Karpukhin2020DensePR} &   78.4  &  85.4 \\
ANCE ~\citep{Xiong2021ApproximateNN} &   81.9  &  87.5 \\
RocketQA ~\citep{Qu2021RocketQAAO} &   82.7   &  88.5  \\
Condenser ~\citep{Gao2021CondenserAP}  &   83.2   &  88.4  \\
PAIR ~\citep{Ren2021PAIRLP} &  83.5  & 89.1   \\
RocketQAv2 ~\citep{Ren2021RocketQAv2AJ} &   83.7  &  89.0 \\
coCondenser~\citep{Gao2022UnsupervisedCA}  &  84.3   &  89.0  \\ \hline
\our{}    &    \textbf{85.2}  &  \textbf{89.7}  \\ \hline
\end{tabular}}
\caption{Results on the test set of Natural Questions (NQ) dataset.
Listed results of SimLM are based on Retriever$_\text{distill}$.}
\label{tab:dpr_result}
\end{table}

For passage retrieval in the open-domain QA setting,
a passage is considered relevant if it contains the correct answer for a given question.
In Table ~\ref{tab:dpr_result},
our model achieves R@20 85.2 and R@100 89.7 on the NQ dataset,
which are comparable to or better than other methods.
For end-to-end evaluation of question answering accuracy,
we will leave it as future work.

\begin{table}[ht]
\centering
\scalebox{0.9}{\begin{tabular}{lc}
\hline
Model   & MRR@10 \\ \hline
BERT$_\text{base}$    &   42.3  \\
ELECTRA$_\text{base}$ &  \textbf{43.7}   \\ \hline
\our{}   &   42.9  \\ \hline
\end{tabular}}
\caption{Re-ranker performance w/ different pre-trained models on the dev set of MS-MARCO passage ranking dataset.}
\label{tab:compare_rerank}
\end{table}

Though SimLM achieves substantial gain for biencoder-based retrieval,
its success for re-ranking is not as remarkable.
In Table ~\ref{tab:compare_rerank},
when used as initialization for re-ranker training,
SimLM outperforms BERT$_\text{base}$ by 0.6\%
but still lags behind ELECTRA$_\text{base}$.

\begin{table}[ht]
\centering
\scalebox{0.9}{\begin{tabular}{@{}lcc@{}}
\hline
               & MRR@10 & R@1k \\ \hline
\textbf{coCondenser}  &  &  \\
BM25 negatives &   35.7  &  97.8   \\
+ mined negatives     &  38.2  & 98.4 \\
+ distillation &  40.2$^*$   &   98.3$^*$  \\ \hline
\textbf{\our{}}           &        &      \\
BM25 negatives (Retriever$_\text{1}$) &  38.0  &  98.3  \\
+ mined negatives (Retriever$_\text{2}$)  &  39.1  &  98.6 \\
+ distillation (Retriever$_\text{distill}$) &  \textbf{41.1} &  \textbf{98.7} \\ \hline
Cross-encoder re-ranker &   43.7  &  98.6  \\ \hline
\end{tabular}}
\caption{Comparison with state-of-the-art dense retriever coCondenser
under various settings on the dev set of MS-MARCO passage ranking dataset.
Results with * are from our reproduction.}
\label{tab:cocon_comparison}
\end{table}

Next,
we zoom in on the impact of each stage in our training pipeline.
In Table ~\ref{tab:cocon_comparison},
we mainly compare with coCondenser ~\citep{Gao2022UnsupervisedCA}.
With BM25 hard negatives only,
we can achieve MRR@10 38.0,
which already matches the performance of many strong models like RocketQA ~\citep{Qu2021RocketQAAO}.
Model-based hard negative mining and re-ranker distillation can bring further gains.
This is consistent with many previous works ~\citep{Xiong2021ApproximateNN, Ren2021RocketQAv2AJ}.
We also tried an additional round of mining hard negatives
but did not observe any meaningful improvement.

Based on the results of Table ~\ref{tab:cocon_comparison},
there are many interesting research directions to pursue.
For example,
how to simplify the training pipeline of dense retrieval systems
while still maintaining competitive performance?
And how to further close the gap between biencoder-based retriever
and cross-encoder based re-ranker?

\section{Analysis}

\subsection{Variants of Pre-training Objectives}

\begin{table*}[ht]
\centering
\scalebox{0.9}{\begin{tabular}{cccccccc}
\hline
       & \our{} & Enc-Dec MLM & Condenser & MLM  & Enc-Dec RTD & AutoEncoder & BERT$_\text{base}$ \\ \hline
MRR@10 & \textbf{38.0} & 37.7  & 36.9 & 36.7 &  36.2 & 32.8 & 33.7 \\ \hline
\end{tabular}}
\caption{Different pre-training objectives.
Reported numbers are MRR@10 on the dev set of MS-MARCO passage ranking.
We finetune the pre-trained models with official BM25 hard negatives.}
\label{tab:objectives}
\end{table*}

Besides our proposed replaced language modeling objective,
we also tried several other pre-training objectives as listed below.
\newline

\noindent
\textbf{Enc-Dec MLM }
uses the same encoder-decoder architecture as in Figure ~\ref{fig:pretrain} but without the generator.
The inputs are randomly masked texts and
the pre-training objective is masked language modeling (MLM) over the masked tokens only.
The mask rate is the same as our method for a fair comparison,
which is 30\% for the encoder and 50\% for the decoder.
In contrast,
RetroMAE ~\citep{Liu2022RetroMAEPR} uses a specialized decoding mechanism
to derive supervision signals from all tokens on the decoder side.

\noindent
\textbf{Condenser }
is a pre-training architecture proposed by ~\citet{Gao2021CondenserAP}.
Here we pre-train Condenser with a 30\% mask rate on the target corpus.

\noindent
\textbf{MLM }
is the same as the original BERT pre-training objective with a 30\% mask rate.

\noindent
\textbf{Enc-Dec RTD }
is the same as our method in Figure ~\ref{fig:pretrain} except that
we use replaced token detection (RTD) ~\citep{Clark2020ELECTRAPT}
as a pre-training task for both the encoder and decoder.
This variant shares some similarities with DiffCSE ~\citep{Chuang2022DiffCSEDC}.
The main difference is that the input for DiffCSE encoder is the original text,
making it a much easier task.
Our preliminary experiments with DiffCSE pre-training
do not result in any improvement.

\noindent
\textbf{AutoEncoder }
attempts to reconstruct the inputs based on the bottleneck representation.
The encoder input is the original text without any mask,
and the decoder input only consists of [MASK] tokens
and [CLS] vector from the encoder.

\noindent
\textbf{BERT$_\text{base}$} just uses off-the-shelf checkpoint published by ~\citet{Devlin2019BERTPO}.
It serves as a baseline to compare against various pre-training objectives.
\newline

The results are summarized in Table ~\ref{tab:objectives}.
Naive auto-encoding only requires memorizing the inputs
and does not need to learn any contextualized features.
As a result,
it becomes the only pre-training objective that underperforms BERT$_\text{base}$.
Condenser is only slightly better than simple MLM pre-training,
which is possibly due to the bypassing effects of the skip connections in Condenser.
Enc-Dec MLM substantially outperforms Enc-Dec RTD,
showing that MLM is a better pre-training task than RTD for retrieval tasks.
This is consistent with the results in Table ~\ref{tab:inconsistent_perf}.
Considering the superior performance of RTD pre-trained models on benchmarks like GLUE,
we believe further research efforts are needed to investigate the reason behind this phenomenon.

\subsection{Effects of Replace Rate}

\begin{table}[ht]
\centering
\begin{tabular}{ccc}
\hline
encoder & decoder & MRR@10 \\ \hline
15\%    & 15\%    &   37.6  \\
15\%    & 30\%    &   37.5  \\
30\%    & 30\%    &   37.9  \\
30\%    & 50\%    &   \textbf{38.0} \\
40\%    & 60\%    &   \textbf{38.0} \\
30\%    & 100\%   &    36.6 \\ \hline
\end{tabular}
\caption{MS-MARCO passage ranking performance w.r.t different token replace rates.
Here the replace rate is the percentage of masked tokens fed to the generator.}
\label{tab:replace_ratio}
\end{table}

In the experiments,
we use fairly large replace rates (30\% for the encoder and 50\% for the decoder).
This is in stark contrast to the mainstream choice of 15\%.
In Table ~\ref{tab:replace_ratio},
we show the results of pre-training with different replace rates.
Our model is quite robust to a wide range of values
with 30\%-40\% encoder replace rate performing slightly better.
Similar findings are also made by ~\citet{Wettig2022ShouldYM}.

One interesting extreme scenario is a 100\% replace rate on the decoder side.
In such a case,
the decoder has no access to any meaningful context.
It needs to predict the original texts solely based on the representation bottleneck.
This task may be too difficult and has negative impacts on the encoder.

\subsection{Effects of Pre-training Steps}

\begin{table*}[ht]
\centering
\scalebox{0.98}{\begin{tabular}{ll}
\hline
\small query    & \small was winnie the pooh a boy   \\ \hline
\small BERT$_\text{base}$ & \small\begin{tabular}[c]{@{}l@{}} \textbf{Rank}: 1, \textbf{Relevant}: \xmark \\ \textbf{Passage}: The little boy who talks to the animals in the Winnie-the-Pooh stories is called Christopher Robin, \\ which is the name of A. A. Milne's real-life son, who was born in 1920. On August 21, 1921, \\ the real-life Christopher Robin Milne received a stuffed bear from Harrods for his first birthday \ldots \end{tabular}   \\ \hline
\small \our{}     & \small\begin{tabular}[c]{@{}l@{}} \textbf{Rank}: 1, \textbf{Relevant}: \cmark \\ \textbf{Passage}: So, it looks like we were lied to our entire childhood! \emph{Winnie the Pooh is not a boy.} SHE is a girl \\ and she's from Canada, not England. Really! In a new picture book called Finding Winnie: \\ The True Story of the World's Most Famous Bear, we learn that Winnie is actually named after \ldots \end{tabular}   \\ \hline \hline
\small query    & \small colorado routing number loveland colorado \\ \hline
\small BERT$_\text{base}$  & \small\begin{tabular}[c]{@{}l@{}} \textbf{Rank}: 1, \textbf{Relevant}: \xmark \\ \textbf{Passage}: Loveland, CO is currently served by one area code which is area code 970. In addition to Loveland, \\ CO area code information read more about area code 970 details and Colorado area codes. \ldots \end{tabular} \\ \hline
\small \our{}   &   \small\begin{tabular}[c]{@{}l@{}} \textbf{Rank}: 2, \textbf{Relevant}: \cmark \\ \textbf{Passage}: \emph{107006787 Routing Transit Number (RTN) for Advantage Bank Main Office located at} \\ \emph{Loveland, Colorado, CO}, 80538, United States, Street Address 1475 NORTH DENVER AVENUE, \\ Telephone Number 970-613-1982 \ldots \end{tabular}  \\ \hline
\end{tabular}}
\caption{Some (cherry-picked) examples from the dev set of MS-MARCO passage ranking dataset.
We show the query, top retrieved passages from different models, and their binary relevance labels.
Relevant text snippets are shown in italic.
More examples are available in the Appendix.}
\label{tab:case_analysis}
\end{table*}

\begin{figure}[ht]
\begin{center}
 \includegraphics[width=0.9\linewidth]{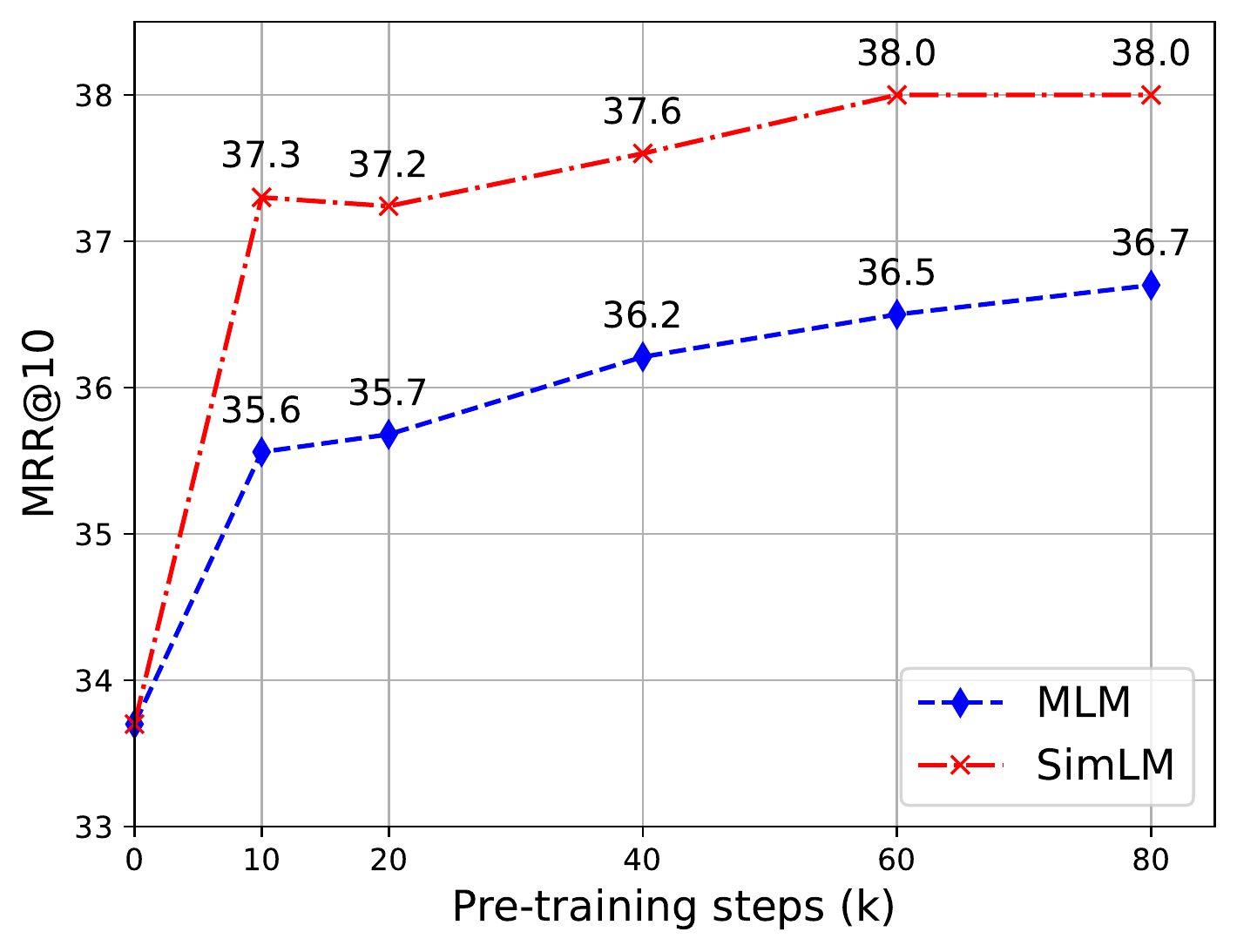}
 \caption{Our pre-training objective converges faster and consistently
 outperforms vanilla masked language model pre-training.
 The y-axis shows the MRR@10 on the dev set of MS-MARCO dataset.}
 \label{fig:converge}
\end{center}
\end{figure}

Since pre-training can be costly in terms of both time and carbon emission,
it is preferred to have an objective that converges fast.
Our proposed method shares two advantages of ELECTRA ~\citep{Clark2020ELECTRAPT}.
First,
the loss is computed over \emph{all} input tokens instead of a small percentage of masked ones.
Second,
the issue of input distribution mismatch is less severe than MLM,
where the [MASK] token is seen during pre-training but not for supervised fine-tuning.
In Figure ~\ref{fig:converge},
our method achieves competitive results with only $10k$ training steps
and converges at $60k$,
while MLM still slowly improves with more steps.

\subsection{On the Choice of Pre-training Corpus}

\begin{table}[ht]
\centering
\scalebox{0.85}{\begin{tabular}{lcccc}
\hline
\multirow{2}{*}{Corpus} & \multicolumn{2}{c}{MS-MARCO} & \multicolumn{2}{c}{NQ} \\
             & MRR@10   & R@1k   & R@20  & R@100  \\ \hline
none   &   33.7   &  95.9   & 82.9  & 88.0  \\
MS-MARCO  &  \textbf{38.0}   &  \textbf{98.3}  &   83.3   &  88.6  \\
Wikipedia  &  36.3  & 97.4  &   \textbf{84.3} &  \textbf{89.3}  \\ \hline
\end{tabular}}
\caption{Fine-tuning performance w.r.t different pre-training corpora.
We use BM25 negatives for MS-MARCO and mined negatives for NQ.
``Wikipedia'' is the target retrieval corpus for NQ dataset.
``none'' use BERT$_\text{base}$ as the foundation model.}
\label{tab:choice_pretrain_corpus}
\end{table}

For a typical retrieval task,
the number of candidate passages is much larger than the number of labeled queries,
and many passages are never seen during training.
Take the NQ dataset as an example,
it has $21M$ candidate passages but only less than $80k$ question-answer pairs for training.
In the experiments,
we directly pre-train on the target corpus.
Such pre-training can be regarded as implicit memorization of the target corpus
in a query-agnostic way.
One evidence to support this argument is that,
as shown in Table ~\ref{tab:objectives},
simple MLM pre-training on target corpus can have large performance gains.

An important research question to ask is:
will there be any benefits of our method when pre-training on non-target corpus?
In Table ~\ref{tab:choice_pretrain_corpus},
the largest performance gains are obtained
when the corpus matches between pre-training and fine-tuning.
If we pre-train on the MS-MARCO corpus and fine-tune on the labeled NQ dataset
or the other way around,
there are still considerable improvements over the baseline.
We hypothesize that
this is due to the model's ability to compress information into a representation bottleneck.
Such ability is beneficial for training robust biencoder-based retrievers.

\subsection{Case Analysis}

To qualitatively understand the gains brought by pre-training,
we show several examples in Table ~\ref{tab:case_analysis}.
The BERT$_\text{base}$ retriever can return passages with high lexical overlap
while missing some subtle but key semantic information.
In the first example,
the retrieved passage by BERT$_\text{base}$ contains keywords like ``boy'', ``Winnie the Pooh'',
but does not answer the question.
In the second example,
there is no routing number in the BERT$_\text{base}$ retrieved passage,
which is the key intent of the query.
Our proposed pre-training can help to learn better semantics to answer such queries.
For more examples,
please check out Table ~\ref{tab:more_cases} in the Appendix.

\section{Conclusion}

This paper proposes a novel pre-training method \our{} for dense passage retrieval.
It follows an encoder-decoder architecture with a representation bottleneck in between.
The encoder learns to compress all the semantic information into a dense vector
and passes it to the decoder to perform well on the replaced language modeling task.
When used as initialization in a dense retriever training pipeline,
our model achieves competitive results on
several large-scale passage retrieval datasets.

For future work,
we would like to increase the model size and the corpus size
to examine the scaling effects.
It is also interesting to explore other pre-training mechanisms
to support unsupervised dense retrieval and multilingual retrieval.

\section*{Limitations}

One limitation of SimLM is that it can not be used as a zero-shot dense retriever,
since the pre-training framework does not have any contrastive objective.
Fine-tuning on labeled data is necessary to get a high-quality model.
On the other hand,
although SimLM pre-training is quite efficient thanks to the replaced language modeling objective,
it still requires extra computational resources to train the model.

\section*{Ethical Considerations}
If the retrieval corpus contains some offensive or biased texts,
they could be exposed to users under certain queries through our dense retriever.
To deal with such risks,
we need to introduce toxic text classifiers or manual inspection to exclude such texts from the corpus.

\bibliography{anthology,custom}
\bibliographystyle{acl_natbib}

\appendix

\section{Details on Table ~\ref{tab:inconsistent_perf}} ~\label{sec:inconsistent}

The numbers for the GLUE benchmark are
from the official leaderboard ~\footnote{~\url{https://gluebenchmark.com/leaderboard}}.
Note that the leaderboard submission from BERT does not use ensemble,
so the comparison is not entirely fair.
However,
this does not change our conclusion that
BERT generally performs worse than RoBERTa and ELECTRA on NLP tasks.
For the MS-MARCO dataset,
we fine-tune all the pre-trained models with BM25 hard negatives only.
For BERT and RoBERTa,
we use the same hyperparameters as discussed in Section ~\ref{sec:impl_details}.
For ELECTRA,
we train for $6$ epochs with a peak learning rate $4 \times 10^{-5}$
since it converges much slower.

\section{Implementation Details} ~\label{sec:more_impl_details}

\begin{table}[ht]
\centering
\scalebox{0.95}{\begin{tabular}{lcc}
\hline
 & MS-MARCO & Wikipedia \\ \hline
\# of passages &  8.8M   &  21M  \\
PLM &  BERT$_\text{base}$  &  BERT$_\text{base}$ \\
batch size &   2048  &  2048   \\
text length &  144   &  144  \\
learning rate &  $3 \times 10^{-4}$  & $3 \times 10^{-4}$ \\
warmup steps & 4000  &  4000  \\
train steps &   $80k$  &  $200k$  \\
encoder replace rate &   30\%  &   30\%  \\
decoder replace rate &   50\%  &   50\%   \\ \hline
\end{tabular}}
\caption{Hyper-parameters for pre-training.
The Wikipedia corpus comes from DPR ~\citep{Karpukhin2020DensePR}
instead of the original one used for BERT pre-training.}
\label{tab:hp_pretrain}
\end{table}

The hyper-parameters for our proposed pre-training
and fine-tuning are listed in Table ~\ref{tab:hp_pretrain} and ~\ref{tab:hp_marco_finetune},
respectively.
For supervised fine-tuning,
One shared encoder is used to encode both the query and passages.
We start with the official BM25 hard negatives in the first training round
and then change to mined hard negatives.
During inference,
given a query,
we use brute force search to rank all the passages for a fair comparison with previous works.
The generator is initialized with the released one by ELECTRA authors ~\footnote{\url{https://huggingface.co/google/electra-base-generator}},
and its parameters are frozen during pre-training.
All the reported results are based on a single run,
we find that the numbers are quite stable with different random seeds.

For fine-tuning on the NQ dataset,
we reuse most hyper-parameters values from MS-MARCO training.
A few exceptions are listed below.
We fine-tune for $20k$ steps with learning rate $5 \times 10^{-6}$.
The maximum length for passage is $192$.
The mined hard negatives come from top-$100$ predictions
that do not contain any correct answer.

\section{Variants of Generators}

In the ELECTRA pre-training,
the generator plays a critical role.
Using either a too strong or too weak generator
hurts the learnability and generalization of the discriminator.

\begin{table}[ht]
\centering
\scalebox{0.95}{\begin{tabular}{lcc}
\hline
generator                  & MRR@10 & R@1k \\ \hline
frozen generator           &  \textbf{38.0}   & 98.3  \\
joint train                &  \textbf{38.0}  &  \textbf{98.4} \\
joint train w/ random init &  37.8  &  \textbf{98.4} \\ \hline
\end{tabular}}
\caption{Variants of generators for SimLM pre-training.
Performances are reported on the dev set of MS-MARCO with BM25 negatives only.}
\label{tab:compare_generator}
\end{table}

We also tried several variants of generators.
In Table ~\ref{tab:compare_generator},
``frozen generator'' keeps the generator parameters unchanged during our pre-training,
``joint train'' also fine-tunes the generator parameters,
and ``joint train w/ random init'' uses randomly initialized generator parameters.
We do not observe any significant performance difference
between these variants.
In our experiments,
we simply use the ``frozen generator'' as it has a faster training speed.

\begin{table*}[ht]
\centering
\begin{tabular}{lccc}
\hline
 & Retriever 1-2 & Re-ranker & Retriever$_\text{distill}$ \\ \hline
learning rate &   $2 \times 10^{-5}$  & $3 \times 10^{-5}$    &  $3 \times 10^{-5}$    \\
PLM & \our{}  & ELECTRA$_\text{base}$  &  \our{} \\
\# of GPUs & 4  & 8  & 4 \\
warmup steps &  1000  &  1000   &  1000  \\
batch size &  64 &  64  &  64 \\
epoch &  3  &  3 &  6 \\
$\tau$ &  0.02  &  n.a.  &  0.02   \\
$\alpha$ & n.a.  & n.a.  & 0.2  \\
negatives depth & 200  & 200  & 200  \\
rerank depth & n.a.  &  200  &  n.a. \\
query length & 32  &  n.a.  &  32 \\
passage length &  144  &  192$^\dagger$  &  144   \\
\# of negatives &  15  &  63  &  23 \\ \hline
\end{tabular}
\caption{Hyper-parameters for supervised fine-tuning on MS-MARCO passage ranking dataset.
$\dagger$: Max length for the concatenation of the query and passage.}
\label{tab:hp_marco_finetune}
\end{table*}

\begin{table*}[ht]
\centering
\scalebox{0.9}{\begin{tabular}{ll}
\hline
\small query    & \small is the keto diet good for kidney disease   \\ \hline
\small BERT$_\text{base}$ & \small\begin{tabular}[c]{@{}l@{}} \textbf{Rank}: 1, \textbf{Relevant}: \xmark \\ \textbf{Passage}: The keto diet (also known as ketogenic diet, low carb diet and LCHF diet) is a low carbohydrate, \\ high fat diet. Maintaining this diet is a great tool for weight loss. More importantly though, \\ according to an increasing number of studies, it helps reduce risk factors for diabetes, heart diseases, stroke \ldots \end{tabular}   \\ \hline
\small \our{}     & \small\begin{tabular}[c]{@{}l@{}} \textbf{Rank}: 1, \textbf{Relevant}: \cmark \\ \textbf{Passage}: 4-Many kidney issues have either a hyperinsulinemic characteristic, an autoimmune characteristic, \\ and or a combination of autoimmunity or hyperinsulinism. A standard, low-ish carb paleo diet can fix most of \\ these issues. \emph{5-For serious kidney damage a low-protein, ketogenic diet can be remarkably therapeutic.} \end{tabular}   \\ \hline \hline
\small query    & \small who announced the european recovery program? \\ \hline
\small BERT$_\text{base}$  & \small\begin{tabular}[c]{@{}l@{}} \textbf{Rank}: 1, \textbf{Relevant}: \xmark \\ \textbf{Passage}: 1 The CEEC submits its report estimating needs and the cost of the European Recovery Program \\ (ERP) over four years. 2  It provides for the establishment of the Organization for European \\ Economic Cooperation (OEEC) to coordinate the program from the European side. 3  February 1948. \end{tabular} \\ \hline
\small \our{}   &   \small\begin{tabular}[c]{@{}l@{}} \textbf{Rank}: 2, \textbf{Relevant}: \cmark \\ \textbf{Passage}: Marshall Plan. Introduction. The Marshall Plan, also known as the European Recovery Program, \\ channeled over \$13 billion to finance the economic recovery \ldots The plan is named for Secretary of State \\ \emph{George C. Marshall, who announced it in a commencement speech at Harvard University on June 5, 1947.} \end{tabular}  \\ \hline \hline
\small query    & \small what is process control equipment \\ \hline
\small BERT$_\text{base}$  & \small\begin{tabular}[c]{@{}l@{}} \textbf{Rank}: 1, \textbf{Relevant}: \xmark \\ \textbf{Passage}: What is process control? Process control is an algorithm that is used in the during the manufacturing \\ process in the industries for the active changing process based on the output of process monitoring. \end{tabular} \\ \hline
\small \our{}   &   \small\begin{tabular}[c]{@{}l@{}} \textbf{Rank}: 1, \textbf{Relevant}: \xmark \\ \textbf{Passage}: Process equipment is equipment used in chemical and materials processing, in facilities \\ like refineries, chemical plants, and wastewater treatment plants. This equipment is usually designed with a \\ specific process or family of processes in mind and can be customized for a particular facility in some cases. \end{tabular}  \\ \hline
\end{tabular}}
\caption{Additional examples from dev set of MS-MARCO passage ranking dataset.}
\label{tab:more_cases}
\end{table*}

\end{document}